\documentclass{aa}

\usepackage{graphicx}
\usepackage{txfonts}

\newcommand{\tablenotea}[1]{\parbox{8.5cm}{\indent \footnotesize{#1}}}
\newcommand{\tablenoteb}[1]{\parbox{10.9cm}{\indent \footnotesize{#1}}}
\newcommand{\tablenotec}[1]{\parbox{16.6cm}{\indent \footnotesize{#1}}}
\newcommand{\tablenoted}[1]{\parbox{15.7cm}{\indent \footnotesize{#1}}}
\newcommand{\tablenotee}[1]{\parbox{7.5cm}{\indent \footnotesize{#1}}}

\newcommand{\jms}{J. Mol. Spectr.}

\newcommand{\pccp}{Phys. Chem. Chem. Phys.}

\newcommand{\science}{Science}

\begin{document}

\title{The rich interstellar reservoir of dinitriles:\\ Detection of malononitrile and maleonitrile in TMC-1\thanks{Based on observations carried out with the Yebes 40m telescope (projects 19A003, 20A014, 20D023, 21A011, 21D005, 22A007, 22B029, and 23A024). The 40m radio telescope at Yebes Observatory is operated by the Spanish Geographic Institute (IGN; Ministerio de Transportes, Movilidad y Agenda Urbana).}}

\titlerunning{Interstellar malononitrile and maleonitrile}
\authorrunning{Ag\'undez et al.}

\author{M.~Ag\'undez\inst{1}, C.~Berm\'udez\inst{2}, C.~Cabezas\inst{1}, G.~Molpeceres\inst{1}, Y.~Endo\inst{3}, N.~Marcelino\inst{4,5}, B.~Tercero\inst{4,5}, J.-C.~Guillemin\inst{6}, P.~de~Vicente\inst{5}, \and J.~Cernicharo\inst{1}}

\institute{
Instituto de F\'isica Fundamental, CSIC, Calle Serrano 123, E-28006 Madrid, Spain\\ \email{marcelino.agundez@csic.es, jose.cernicharo@csic.es} 
\and
Departamento de Qu\'imica F\'isica y Qu\'imica Inorg\'anica, Facultad de Ciencias $-$ I. U. CINQUIMA, Universidad de Valladolid, Valladolid 47011, Spain
\and
Department of Applied Chemistry, Science Building II, National Yang Ming Chiao Tung University, 1001 Ta-Hsueh Rd., Hsinchu 300098, Taiwan
\and
Observatorio Astron\'omico Nacional, IGN, Calle Alfonso XII 3, E-28014 Madrid, Spain 
\and
Observatorio de Yebes, IGN, Cerro de la Palera s/n, E-19141 Yebes, Guadalajara, Spain
\and
Univ Rennes, Ecole Nationale Sup\'erieure de Chimie de Rennes, CNRS, ISCR – UMR6226, F-35000 Rennes, France
}

\date{Received; accepted}

 
\abstract
{While the nitrile group is by far the most prevalent one among interstellar molecules, the existence of interstellar dinitriles (molecules containing two $-$CN groups) has recently been proven. Here we report the discovery of two new dinitriles in the cold dense cloud \mbox{TMC-1}. These newly identified species are malononitrile, CH$_2$(CN)$_2$, and maleonitrile, the $Z$ isomer of NC$-$CH$=$CH$-$CN, which can be seen as the result of substituting two H atoms with two $-$CN groups in methane and ethylene, respectively. These two molecules were detected using data from the ongoing QUIJOTE line survey of \mbox{TMC-1} that is being carried out with the Yebes\,40m telescope. We derive column densities of 1.8\,$\times$\,10$^{11}$ cm$^{-2}$ and 5.1\,$\times$\,10$^{10}$ cm$^{-2}$ for malononitrile and maleonitrile, respectively. This means that they are eight and three times less abundant than HCC$-$CH$_2$$-$CN and ($E$)-HCC$-$CH$=$CH$-$CN, respectively, which are analog molecules detected in \mbox{TMC-1} in which one $-$CN group is converted into a $-$CCH group. This is in line with previous findings in which $-$CCH derivatives are more abundant than the $-$CN counterparts in \mbox{TMC-1}. We examined the potential chemical pathways to these two dinitriles, and we find that while maleonitrile can be efficiently formed through the reaction of CN with CH$_2$CHCN, the formation of malononitrile is not clear because the neutral-neutral reactions that could potentially form it are not feasible under the physical conditions of \mbox{TMC-1}.}


\maketitle

\section{Introduction}

The nitrile group ($-$C$\equiv$N) is the most prevalent functional group among interstellar molecules. It is more common than, for example, the hydroxyl group ($-$OH), reflecting the fact that interstellar chemistry is largely organic in nature. Being a highly polar group, they are very convenient for probing nonpolar molecules such as benzene, which is indirectly probed through C$_6$H$_5$CN \citep{McGuire2018,Cooke2020}. The high occurrence of nitriles in interstellar clouds is also interesting because this group plays a key role in prebiotic chemistry. In fact, several molecules of prebiotic interest detected in the interstellar medium, such as glycolonitrile \citep{Zeng2019}, contain this group.

In recent years, it has been shown that molecules containing two $-$C$\equiv$N groups are also present in interstellar space. The simplest such molecule is cyanogen (NCCN), whose protonated form was detected in cold clouds \citep{Agundez2015}. More recently, a metastable isomer of cyanogen known as isocyanogen (CNCN) was also detected \citep{Agundez2018}, and the member of the family of dicyanopolyynes N$\equiv$C$-$C$\equiv$C$-$C$\equiv$N was probed through its protonated form \citep{Agundez2023a}. Dicyanopolyynes are very stable molecules, and their presence in interstellar space was suggested more than 20 years ago \citep{Kolos2000,Petrie2003}. To our knowledge, no other molecule containing two nitrile groups has been detected in the interstellar medium, although very recently \cite{SanAndres2024} detected a molecule with a skeleton similar to that of isocyanogen, H$_2$CNCN.

Here we report the interstellar detection of malononitrile, CH$_2$(CN)$_2$, and maleonitrile, the $Z$ isomer of NC$-$CH$=$CH$-$CN, toward the cold dense cloud \mbox{TMC-1}. Malononitrile and maleonitrile can be seen as the result of substituting two H atoms with two $-$CN groups in methane and ethylene. Malononitrile is an interesting molecule because it may have played a key role in the pre-RNA world in the early Earth as a precursor of purines and nucleosides \citep{MenorSalvan2020} and may be present in the atmosphere of Titan \citep{Ramal-Olmedo2023}. The present work demonstrates that malononitrile and maleonitrile are readily formed in cold interstellar clouds, in much earlier evolutionary stages.


\section{Astronomical observations}

\begin{figure}
\centering
\includegraphics[angle=0,width=0.96\columnwidth]{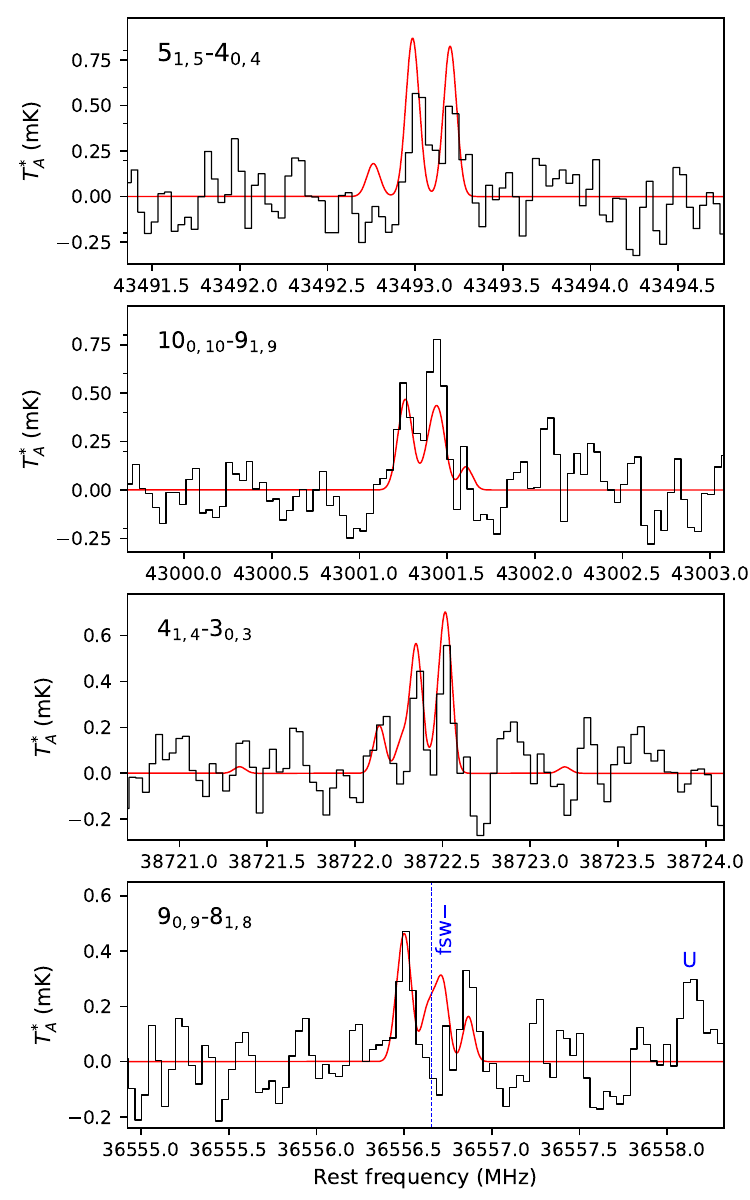}
\caption{Rotational lines of malononitrile, CH$_2$(CN)$_2$, observed toward \mbox{TMC-1} (see the line parameters in Table\,\ref{table:lines}). The position of negative artifacts caused by the frequency-switching technique are indicated by a dashed blue line with the label "fsw$-$."} \label{fig:lines_malononitrile}
\end{figure}

The astronomical observations reported here are from the ongoing Yebes\,40m Q-band line survey of \mbox{TMC-1,} QUIJOTE (Q-band Ultrasensitive Inspection Journey to the Obscure TMC-1 Environment; \citealt{Cernicharo2021a}). Briefly, QUIJOTE consists of a line survey in the Q band (31.0-50.3 GHz) at the position of the cyanopolyyne peak of \mbox{TMC-1} ($\alpha_{J2000}=4^{\rm h} 41^{\rm  m} 41.9^{\rm s}$ and $\delta_{J2000}=+25^\circ 41' 27.0''$). The observations are carried out in the frequency-switching observing mode, with a frequency throw of 8 or 10 MHz, using a 7 mm receiver connected to a fast Fourier transform spectrometer; this allows the full Q band to be covered in one shot with a spectral resolution of 38.15 kHz \citep{Tercero2021}. The intensity scale is the antenna temperature, $T_A^*$, which has an estimated uncertainty due to calibration of 10\,\%. The main beam brightness temperature, $T_{\rm mb}$, can be obtained by dividing $T_A^*$ by $B_{\rm eff}$/$F_{\rm eff}$, where the main beam and forward efficiencies are given by $B_{\rm eff}$\,=\,0.797\,exp[$-$($\nu$(GHz)/71.1)$^2$] and $F_{\rm eff}$\,=\,0.97. The half power beam width (HPBW) can be approximated as HPBW($''$) = 1763/$\nu$(GHz). We adopted the latest QUIJOTE dataset, as used in previous recent discoveries of molecules, such as HCNS \citep{Cernicharo2024a}, CH$_3$CH$_2$CCH \citep{Cernicharo2024b}, NCCHCS \citep{Cabezas2024a}, HC$_3$N$^+$ \citep{Cabezas2024b}, HC$_5$N$^+$, and HC$_7$N$^+$ \citep{Cernicharo2024c}. The total on-source telescope time invested was 1202 h, of which 465 h correspond to a frequency throw of 8 MHz and 737 h to a throw of 10 MHz. The $T_A^*$ rms noise level varies between 0.07 mK at 32 GHz and 0.2 mK at 49.5 GHz. All data were analyzed using the GILDAS software\footnote{https://www.iram.fr/IRAMFR/GILDAS/}.

\begin{figure*}
\centering
\includegraphics[angle=0,width=0.97\textwidth]{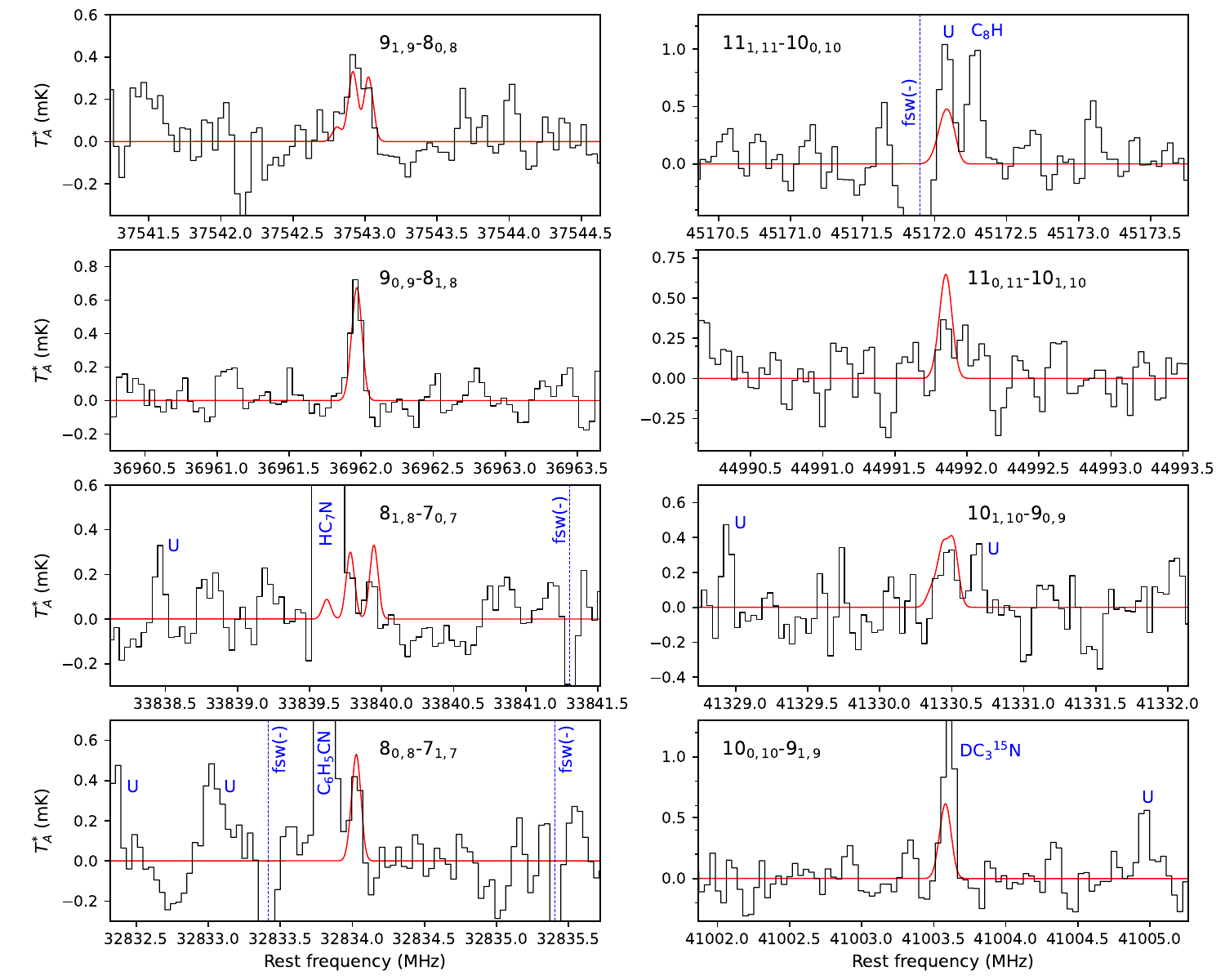}
\caption{Rotational lines of maleonitrile, ($Z$)-NC$-$CH$=$CH$-$CN, observed toward \mbox{TMC-1} (see the line parameters in Table\,\ref{table:lines}). The position of negative artifacts caused by the frequency-switching technique are indicated by a dashed blue line with the label "fsw$-$."} \label{fig:lines_maleonitrile}
\end{figure*}

\begin{table}
\small
\caption{Observed line parameters in \mbox{TMC-1}.}
\label{table:lines}
\centering
\begin{tabular}{crrc}
\hline \hline
Transition & $\nu_{\rm calc}$ (MHz)\,$^a$ & $E_{up}$ (K) & $\int T_A^* dv$ (mK km s$^{-1}$)\,$^b$ \\
\hline
\\
\multicolumn{4}{c}{malononitrile, CH$_2$(CN)$_2$} \\
\hline
9$_{0,9}$-8$_{1,8}$ & 36556.623 & 11.9 & 0.63\,$\pm$\,0.22\,$^c$ \\
4$_{1,4}$-3$_{0,3}$ & 38722.403 & 3.5 & 0.79\,$\pm$\,0.22 \\
10$_{0,10}$-9$_{1,9}$ & 43001.377 & 14.6 & 1.07\,$\pm$\,0.24 \\
5$_{1,5}$-4$_{0,4}$ & 43493.060 & 4.8 & 0.91\,$\pm$\,0.23 \\
\\
\multicolumn{4}{c}{maleonitrile, ($Z$)-NC$-$CH$=$CH$-$CN} \\
\hline
8$_{0,8}$-7$_{1,7}$     & 32834.020 &  7.5 & 0.37\,$\pm$\,0.24 \\
8$_{1,8}$-7$_{0,7}$     & 33839.817 &  7.5 & $-$\,$^d$ \\
9$_{0,9}$-8$_{1,8}$     & 36961.959 &  9.3 & 0.54\,$\pm$\,0.06 \\
9$_{1,9}$-8$_{0,8}$     & 37542.932 &  9.3 & 0.50\,$\pm$\,0.12 \\
10$_{0,10}$-9$_{1,9}$   & 41003.566 & 11.3 & $-$\,$^e$ \\
10$_{1,10}$-0$_{0,9}$   & 41330.441 & 11.3 & 0.33\,$\pm$\,0.11 \\
11$_{0,11}$-10$_{1,10}$ & 44991.836 & 13.4 & 0.22\,$\pm$\,0.11 \\
11$_{1,11}$-10$_{0,10}$ & 45172.060 & 13.4 & $-$\,$^e$ \\
\hline
\end{tabular}
\tablenotea{\\
$^a$\,Calculated frequency of the centroid of all hyperfine components.
$^b$\,Sum of the velocity-integrated areas of the various components that can be resolved (typically two or three; see Figs.\,\ref{fig:lines_malononitrile} and\,\ref{fig:lines_maleonitrile}), each fitted with a Gaussian.\\
$^c$\,Lower limit because intensity is reduced by the overlap of a frequency-switching  negative artifact.\\
$^d$\,Line is only marginally detected. No fit is attempted.\\
$^e$\,Line is blended with a strong line. No fit is attempted.
}
\end{table}

\section{Results and discussion}

\subsection{Detection and abundance} \label{sec:detection}

The rotational spectrum of malononitrile, CH$_2$(CN)$_2$, has been measured in the laboratory by \cite{Hirota1960}, \cite{Cook1974}, \cite{Burie1982}, and \cite{Cox1985}, and more recently by \cite{Motiyenko2019}. The only nonzero dipole moment lies along its $b$ axis and has been measured to be 3.735\,$\pm$\,0.017 D \citep{Hirota1960}. Spin statistics arising from the symmetry of the molecule (essentially the presence of two equivalent H nuclei and two equivalent N nuclei) need to be taken into account to establish the statistical weights of the rotational levels, which results in {reduced weights of 5 and 7} for levels with even and odd ($K_a$\,+\,$K_c$), respectively. In addition, the rotational levels split into hyperfine sublevels due to the nuclear quadrupole of the two N nuclei. We used the SPCAT program \citep{Pickett1991} to predict the line frequencies from the spectroscopic parameters determined by \cite{Motiyenko2019}. The rotational lines of malononitrile in the Q-band spectrum of \mbox{TMC-1} split into several hyperfine components that can be resolved at our spectral resolution of 38.15 kHz. We note that predicted frequencies have errors of $\sim$\,5 kHz (i.e., much lower than the spectral resolution). The lines that are predicted to be most intense at 9 K, which is the gas kinetic temperature in \mbox{TMC-1} \citep{Agundez2023b}, are shown in Fig.\,\ref{fig:lines_malononitrile}. It is seen that the lines are relatively weak, with antenna temperatures below 1 mK, although the fact that the rotational lines split into resolvable components aids in the identification. There are a few other lines of malononitrile that are predicted to have comparable intensities, but either they overlap with negative artifacts caused by the frequency-switching technique, as occurs for the 3$_{1,3}$-2$_{0,2}$ line, or they lie in excessively noisy regions, such as a handful of lines at frequencies above 45 GHz. The non-negligible errors in the line intensities and the relatively narrow range of upper level energies covered (see Table\,\ref{table:lines}) prevent us from confidently determining a rotational temperature. We thus adopted a rotational temperature equal to the kinetic temperature of the gas, 9 K, which is consistent with the observed relative line intensities, and assumed that malononitrile has a circular emission distribution with a diameter of 80$''$, similar to other N-bearing organic molecules such as C$_6$H$_5$CN \citep{Cernicharo2023}. We derive a column density for malononitrile of 1.8\,$\times$\,10$^{11}$ cm$^{-2}$, which is consistent with the upper limit of 5\,$\times$\,10$^{12}$ cm$^{-2}$ derived by \cite{Matthews1986}, although it is somewhat higher than the more recent upper limit of 7.3\,$\times$\,10$^{10}$ cm$^{-2}$ derived by \cite{Cordiner2017}.

In the case of maleonitrile, ($Z$)-NC$-$CH$=$CH$-$CN, the rotational spectrum was measured in the 5-15 GHz region by \cite{Halter2001}, and more recently in the 75-116 GHz range using the GACELA (GAs CEll for Laboratory Astrophysics; \citealt{Cernicharo2019}) setup (see Appendix\,\ref{app:maleonitrile} for a detailed account of these measurements). Maleonitrile only has a nonzero dipole moment along its $b$ axis, measured to be 5.32\,$\pm$\,0.06 D \citep{Halter2001}. As in the case of malononitrile, the rotational levels are affected by spin statistics, with weights of 5 and 7 for even and odd ($K_a$\,+\,$K_c$), respectively, and by hyperfine structure due to the nuclear quadrupole of the two N nuclei (see more details in Appendix\,\ref{app:maleonitrile}). The lines of maleonitrile detected in \mbox{TMC-1} are shown in Fig.\,\ref{fig:lines_maleonitrile}. The hyperfine splitting is very small for $K_a$\,=\,0 lines but can be resolved for $K_a$\,$\geq$\,1 lines. The observed lines have errors in their predicted frequencies of $\sim$\,20 kHz, which still allows for a confident assignment of the lines. The 12$_{0,12}$-11$_{1,11}$ and 12$_{1,12}$-11$_{0,11}$ lines, lying at around 49 GHz, are predicted with similar intensities to those of the analog lower-$J$ lines, but they are not detected because the noise is significantly higher in that region of the spectrum. For maleonitrile, we also adopted a rotational temperature of 9 K, because it cannot be confidently constrained, and an emission size diameter of 80$''$ to derive a column density of 5.1\,$\times$\,10$^{10}$ cm$^{-2}$.

We also searched for other molecules that result from the substitution of two H atoms with $-$CN, $-$NC, and/ or $-$CCH groups in methane and ethylene. Malononitrile has two isomers that arise when one or two of the cyano ($-$CN) groups turn into isocyano ($-$NC): isocyanoacetonitrile (NC$-$CH$_2$$-$NC) and diisocyanomethane (CN$-$CH$_2$$-$NC). The rotational spectrum is known for the two molecules \citep{Motiyenko2012,Motiyenko2019}, but neither of them are detected in \mbox{TMC-1}. The 3\,$\sigma$ upper limits to their column densities (see Table\,\ref{table:column_densities}) imply that they are at least four times less abundant than malononitrile. A molecule related to malononitrile in which one of the nitrile groups is exchanged with an ethynyl group, HCC$-$CH$_2$$-$CN, has also been detected in \mbox{TMC-1} (\citealt{McGuire2020,Marcelino2021}; see also our Appendix\,\ref{app:line-by-line}). This molecule is eight times more abundant than malononitrile (see Table\,\ref{table:column_densities}). This is in line with $-$CCH derivatives being more abundant than $-$CN derivatives, for example CH$_2$CCHCCH/CH$_2$CCHCN\,=\,4.4 \citep{Marcelino2021,Cernicharo2021b}, C$_6$H$_5$CCH/C$_6$H$_5$CN\,=\,2.5 \citep{Cernicharo2021c,Loru2023}, and CH$_2$CHCCH/CH$_2$CHCN\,=\,1.8 \citep{Cernicharo2021d}, which reflects the fact that CCH is ten times more abundant than CN in \mbox{TMC-1} \citep{Pratap1997}.


\begin{table}
\small
\caption{Column densities in \mbox{TMC-1}.}
\label{table:column_densities}
\centering
\begin{tabular}{l@{\hspace{0.05cm}}rl@{\hspace{0.1cm}}r}
\hline \hline
Molecule & $N$ (cm$^{-2}$) & Molecule & $N$ (cm$^{-2}$) \\
\hline
CH$_2$(CN)$_2$ & 1.8\,$\times$\,10$^{11}$ & ($Z$)-NC$-$CH$=$CH$-$CN & 5.1\,$\times$\,10$^{10}$ \\
NC$-$CH$_2$$-$NC & $<$5.1\,$\times$\,10$^{10}$ & CH$_2$$=$C(CN)$_2$ & $<$2.5\,$\times$\,10$^{10}$ \\
CN$-$CH$_2$$-$NC & $<$5.5\,$\times$\,10$^{10}$ & (E)-HCC$-$CH$=$CH$-$CN & 1.7\,$\times$\,10$^{11}$ \\ 
HCC$-$CH$_2$$-$CN & 1.4\,$\times$\,10$^{12}$ & ($Z$)-HCC$-$CH$=$CH$-$CN & $<$7.1\,$\times$\,10$^{10}$ \\ 
\hline
\end{tabular}
\end{table}

In the case of ethylene, there are three isomers resulting from the substitution of two H atoms with $-$CN groups. The $E$ isomer of NC$-$CH$=$CH$-$CN (fumaronitrile) is the most stable one and is nonpolar, the $Z$ isomer (maleonitrile) lies 263 K above in energy, and the third one, CH$_2$$=$C(CN)$_2$ (1,1-dicyanoethylene), which lies 1807 K above fumaronitrile, is not detected in \mbox{TMC-1} at the current sensitivity level of our data (see Table\,\ref{table:column_densities}). The rotational spectrum of CH$_2$$=$C(CN)$_2$ was measured by \cite{Tan1978} in the 19-37 GHz range, although the hyperfine structure due to the two N atoms was not resolved. Since the hyperfine structure is important for observing this species in \mbox{TMC-1}, we measured the rotational spectrum of CH$_2$$=$C(CN)$_2$ at high spectral resolution to determine its nuclear quadrupole coupling constants (see Appendix\,\ref{app:ch2c_cn_2}). The substitution of two H atoms with one cyano and one ethynyl group in ethylene yields three different isomers. The $E$ isomer of HCC$-$CH$=$CH$-$CN has been detected in \mbox{TMC-1} using line stack techniques (\citealt{Lee2021}; see our Appendix\,\ref{app:line-by-line} for a solid line-by-line detection), and it is three times more abundant than maleonitrile, in line with the higher abundance of $-$CCH derivatives compared to $-$CN ones mentioned above. The $Z$ isomer of HCC$-$CH$=$CH$-$CN has been spectroscopically characterized in the laboratory by \cite{Halter2001}, although it is not detected, implying that it is less abundant than its $E$ counterpart in \mbox{TMC-1} (see Table\,\ref{table:column_densities}). For the third isomer, CH$_2$$=$C(CN)CCH, the rotational spectrum has not been measured. There is another isomer with the same molecular formula, C$_5$H$_3$N, which is vinylcyanoacetylene or 4-penten-2-ynenitrile (CH$_2$$=$CH$-$C$_3$N). This species has been detected in \mbox{TMC-1} \citep{Lee2021}, although it does not belong to the same family of doubly substituted $-$CN/$-$CCH derivatives of ethylene.


  
\subsection{Chemistry} \label{sec:chemistry}

The synthesis of malononitrile in cold interstellar clouds, such as \mbox{TMC-1}, is not clear. This molecule is not included in the chemical kinetics databases KIDA \citep{Wakelam2015} or UMIST \citep{Millar2024}. The recent theoretical study by \cite{Ramal-Olmedo2023} shed some light on the potential gas-phase neutral-neutral reactions that could form this species in cold media. The most obvious way of forming malononitrile is the reaction between CN and CH$_3$CN, where one hydrogen atom of CH$_3$CN is replaced by the CN radical. The rate coefficient of the reaction CN + CH$_3$CN has been measured;  above room temperature it shows the typical Arrhenius behavior \citep{Zabarnick1989}, while at low temperatures it experiences a dramatic enhancement that is attributed to the three-body stabilization of the van der Waals complex CH$_3$CN\,$\cdot$$\cdot$$\cdot$\,CN \citep{Sleiman2016}. Indeed, the calculations by \cite{Ramal-Olmedo2023} show that the path leading to CH$_2$(CN)$_2$ has a barrier, which implies that malononitrile cannot form this way at low temperatures. According to \cite{Ramal-Olmedo2023}, other bimolecular reactions that could produce malononitrile are endothermic (e.g., CH$_2$CN + HCN, CH$_2$CN + HNC, and CH$_3$ + NCCN), have barriers (e.g., N + CH$_2$CHCN, NH$_2$ + HC$_3$N, H$_2$CN + HCCN, and CN + CH$_3$CH$_2$CN), or preferentially lead to the undetected isomer isocyanoacetonitrile (NC$-$CH$_2$$-$NC), as in the case of H$_2$CNH + C$_2$N. The related molecule HCC$-$CH$_2$$-$CN is thought to form through the reaction CN + CH$_2$CCH$_2$, as discussed by \cite{Marcelino2021}. An analog route that could yield malononitrile would be the reaction CN + CH$_2$CNH, although it is unlikely to be the main formation route to malononitrile because ketenimine (CH$_2$CNH) is not detected in \mbox{TMC-1} and thus it is likely not abundant enough to be a precursor of malononitrile.

None of the aforementioned bimolecular neutral-neutral reactions seem to provide an efficient formation path to malononitrile. An alternative would be to consider radiative associations. It has recently been shown that these reactions can be fast at low temperatures, even for relatively small systems such as CH$_3$ + CH$_3$O \citep{Balucani2015,Tennis2021}. Let us first consider the formation of malononitrile through the radiative association between CN and CH$_2$CN. This reaction has recently been studied through quantum chemical calculations \citep{Ramal-Olmedo2023,Vieira2023}, and it has been found that it is exothermic and barrier-less. Moreover, \cite{Vieira2023} calculated a relatively high capture rate coefficient of close to 10$^{-10}$ cm$^3$ s$^{-1}$ in the 50-300 K temperature range using capture theory. For the route presented in \citet{Vieira2023} to be applicable to \mbox{TMC-1}, the nascent malononitrile molecule needs to be stabilized radiatively. The probability of this relaxation depends on several factors, the most important being the absence of competitive bimolecular exothermic channels. The most obvious bimolecular channel involves the formation of CH(CN)$_2$ + H. We carried out quantum chemical calculations at the $\omega$B97M-D4/ma-def2-TZVP level \citep{Najibi2020,Zheng2011} using ORCA 5.0.4 \citep{Neese2020} and find that this bimolecular channel is exothermic by 24.8 kcal mol$^{-1}$. In addition, the calculated vibrational spectrum of malononitrile does not show intense modes, in contrast with the abovementioned case of CH$_3$OCH$_3$ \citep{Tennis2021}. Consequently, the Einstein coefficients for radiative relaxation are likely small. While we cannot completely rule out the radiative association channel, our calculations suggest that it should be negligible. Another reaction of radiative association that could directly form malononitrile is CH$_2$ + NCCN. The radical methylene is expected to be abundant in \mbox{TMC-1}, around 10$^{-7}$ relative to H$_2$ according to our chemical modeling calculations (see below), while cyanogen is also expected to be fairly abundant, in the range 10$^{-9}$-10$^{-7}$ relative to H$_2$, according to estimations based on the detection of NCCNH$^+$ and CNCN \citep{Agundez2015,Agundez2018}. In addition to the difficulties for the radiative association discussed above, the CH$_2$ + NCCN reaction has an additional handicap. The electronic ground state of CH$_2$ is $^{3}$B$_{1}$, and  as such the formation of CH$_2$(CN)$_2$ from CH$_2$ + NCCN is spin-forbidden. Therefore, the reaction in the triplet channel has a sizable barrier of 3.9 kcal mol$^{-1}$, calculated at the same level of theory as above. Since the most obvious neutral-neutral reactions are discarded, ion-neutral reactions would need to be considered. We however postpone their study to a future work.

We now discuss the formation mechanism of the other dinitrile reported here, maleonitrile. One of the obvious formation reactions is
\begin{subequations} \label{reac:cn+ch2chcn}
\begin{align}
\rm CN + CH_2CHCN & \rightarrow \rm E{\textit -}NC{-}CH{=}CH{-}CN + H,\\ 
                                    & \rightarrow \rm Z{\textit -}NC{-}CH{=}CH{-}CN + H,\\ 
                                   & \rightarrow \rm CH_2C(CN)_2 + H, 
\end{align}
\end{subequations}
where we consider that the three different isomers can be formed. Fortunately, this reaction was recently studied by \cite{Marchione2022} using crossed molecular beams and electronic structure calculations. The authors find that the reaction is barrier-less and fast at low temperatures, with a total rate coefficient of 2.5\,$\times$\,10$^{-11}$ cm$^3$ s$^{-1}$, and that the three isomers can be formed, although the $E$ and $Z$ isomers of NC$-$CH$=$CH$-$CN are favored over CH$_2$$=$C(CN)$_2$. The calculated branching ratios at 10 K for channels (1a), (1b), and (1c) are 0.65, 0.34, and 0.0015. If this reaction is the main route to maleonitrile, the calculated yields indicate that 1,1-dicyanoethylene would be 200 times less abundant than maleonitrile in \mbox{TMC-1}, which is in line with the observational upper limit derived here (see Table\,\ref{table:column_densities}). It would also mean that the nonpolar $E$ isomer of NC$-$CH$=$CH$-$CN is also present in \mbox{TMC-1} with an abundance twice that of maleonitrile.

To evaluate whether the reaction CN + CH$_2$CHCN can provide an efficient route to maleonitrile in \mbox{TMC-1,} we carried out chemical modeling calculations. We adopted a standard model of a cold dense cloud \citep{Agundez2013} and the latest release of the UMIST Database for Astrochemistry \citep{Millar2024}. We included maleonitrile as a new species, assuming that it is formed through reaction (1b) with a rate coefficient of 8.69\,$\times$\,10$^{-12}$ cm$^3$ s$^{-1}$, as calculated by \cite{Marchione2022}, and is destroyed through reactions with abundant cations such as HCO$^+$, H$_3$O$^+$, and H$_3^+$. The calculated peak abundance of maleonitrile in the chemical model lies three to four orders of magnitude below the observed value, but this is mainly because the model underestimates by a similar amount the abundance of the precursor CH$_2$CHCN, which is observed to be 6.5\,$\times$\,10$^{-10}$ relative to H$_2$ \citep{Cernicharo2021d}. Although the chemistry of maleonitrile is modeled in a very crude way, our chemical modeling calculations suggest that the reaction CN + CH$_2$CHCN could indeed be the main source of maleonitrile in \mbox{TMC-1}.

\section{Conclusions}

We have presented the first detection in space of two new dinitriles: malononitrile and maleonitrile. The column densities derived in the cold dense cloud \mbox{TMC-1} are 1.8\,$\times$\,10$^{11}$ cm$^{-2}$ and 5.1\,$\times$\,10$^{10}$ cm$^{-2}$, respectively. These two molecules, which can be seen as the result of the substitution of two H atoms with two $-$CN groups in methane and ethylene, respectively, are eight and three times less abundant than their corresponding counterparts in which one of the $-$CN groups is replaced by a $-$CCH group, namely HCC$-$CH$_2$$-$CN and ($E$)-HCC$-$CH$=$CH$-$CN, respectively, molecules that are already known to be present in \mbox{TMC-1}. We are unable to identify a feasible chemical route to malononitrile, while in the case of maleonitrile, the reaction between CN and CH$_2$CHCN arises as the most likely formation route.

\begin{acknowledgements}

We acknowledge funding support from Spanish Ministerio de Ciencia e Innovaci\'on through grants PID2019-106110GB-I00, PID2019-107115GB-C21, PID2019-106235GB-I00, and PID2022-136525NA-I00. We thank the Consejo Superior de Investigaciones Cient\'ificas (CSIC) for funding through project PIE 202250I097. C.B. thanks Universidad de Valladolid for her "Mar\'ia Zambrano" grant CONVREC-2021-317. G.M. acknowledges the support of the grant RYC2022-035442-I funded by MCIN/AEI/10.13039/501100011033 and ESF+. J.-C.G. thanks the french national program PCMI (Physics and Chemistry of the Interstellar Medium) (INSU-CNRS) for a grant. We acknowledge the referee, R. A. Motiyenko, for a careful reading of the manuscript and a constructive report.

\end{acknowledgements}

\clearpage
\onecolumn

\begin{appendix}

\section{Rotational spectroscopy of maleonitrile} \label{app:maleonitrile}

\subsection{Experimental methods}

The experimental measurements were carried out using the broadband high resolution rotational spectrometer GACELA, installed at the Yebes Observatory within the context of the ERC Synergy project NANOCOSMOS. The spectrometer is equipped with radio receivers similar to those used at the Yebes\,40m telescope. The receivers are equipped with 16\,$\times$\,2.5 GHz fast Fourier transform spectrometers with a spectral resolution of 38.15 kHz, allowing the observation of rotational transitions in the Q (31.5-50 GHz) and W (72-116.5 GHz) bands. A detailed description of the system is given by \cite{Cernicharo2019}.

The maleonitrile sample was synthesized following \cite{Linstead1952}. We placed the sample into a Pyrex$^{TM}$ vacuum Schlenk that is directly connected to the spectrometer cell, which consists of a stainless steel cylinder of 890 mm length and 490 mm diameter. The vapor pressure of maleonitrile at 308 K was sufficient to measure the rotational spectrum in a flow mode. Prior to the sample introduction, the pressure inside the vacuum chamber was 2.0\,$\times$\,10$^{-4}$ mbar. During the experiment the pressure was kept at 3.0\,$\times$\,10$^{-3}$ mbar. Higher pressures produce undesirable line broadenings. We used the frequency-switching technique with two different frequency throws of 25 MHz and 37 MHz. This observing mode has been previously confirmed as the most suitable because the lines are observed twice and the noise improves by a factor of a square root of two. A total of 12 h of effective observation time was accumulated to obtain the spectra at each frequency throw (see Fig.\,\ref{fig:spectrum}).

\subsection{Analysis of the rotational spectrum}

\begin{figure*}[hb!]
\centering
\includegraphics[angle=0,width=0.99\textwidth]{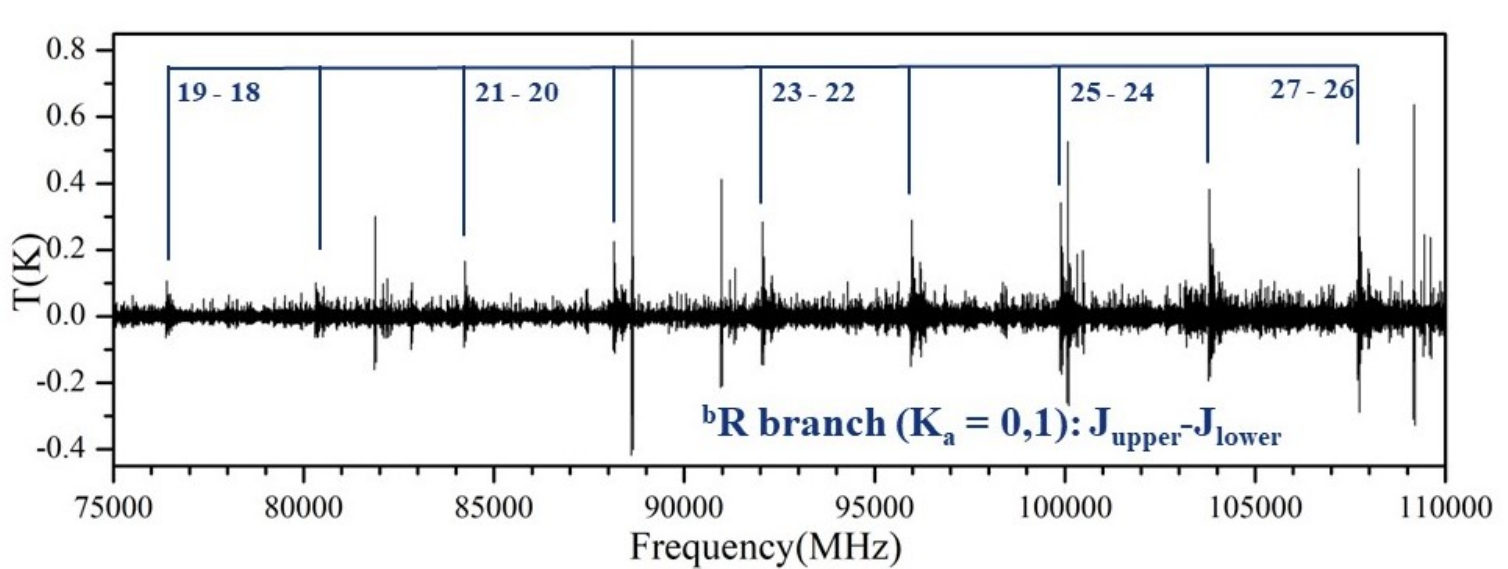}
\caption{Rotational spectrum of maleonitrile obtained with the GACELA setup.}
\label{fig:spectrum}
\end{figure*}

The rotational spectrum of maleonitrile obtained from the GACELA setup clearly shows several groups of intense transitions nearly equally spaced by $\sim$\,3.9 GHz (see Fig.\,\ref{fig:spectrum}). The predictions for the ground vibrational state based on the previous fit using microwave frequencies \citep{Halter2001} show a similar pattern. An iterative process of assignment, fit, and prediction was performed using the CALPGM suite programs SPFIT/SPCAT \citep{Pickett1991} combined with the PROSPE interfaces ASCP/SVIEW \citep{Prospe2001Kisiel}. A total of 108 $b$-type $R$-branch ($J$ values from 18 to 29 and $K_a$ from 0 to 8) and 430 $b$-type $Q$-branch ($J$ values from 10 to 69 and $K_a$ from 1 to 13) transitions were finally observed for maleonitrile.

\begin{table*}
\small
\centering
\caption[]{Experimental spectroscopic parameters for the ground vibrational state of maleonitrile.}
\scalebox{1}{
\label{ground_constants}}
\begin{tabular}{lcccc}
\hline
\hline
Parameter / Unit   & This work\,$^a$ & This work HFS\,$^b$ & \citet{Halter2001} \\
\hline
$A$ / MHz                      & 7389.5215(14)   & 7389.51061(35) & 7389.5090(7)  \\
$B$ / MHz                      & 2672.59327(85)  & 2672.59436(11) & 2672.5944(7)  \\
$C$ / MHz                      & 1959.52286(91)  & 1959.52222(12) & 1959.5221(12) \\
$\Delta_J$ / kHz               &    3.2566(16)   &    3.25637(37) &    3.258(7)   \\
$\Delta_{JK}$ / kHz            &  $-$21.6019(19)   &  $-$21.59241(82) &  $-$21.51(23)   \\
$\Delta_K$ / kHz               &   46.324(22)    &   46.1525(56)  &   45.7(4)     \\
$\delta_J$ / kHz               &    1.18701(14)  &    1.187996(38)&    1.190(3)   \\
$\delta_K$ / kHz               &    3.7917(28)   &    3.7715(13)  &    3.7(5)     \\
$\Phi_J$ / Hz                  &    0.0176(10)   &    0.01710(33) &               \\
$\Phi_{JK}$ / Hz               &   $-$0.0532(39)   &   $-$0.0271(25)  &               \\
$\Phi_{KJ}$ / Hz               &   $-$0.487(14)    &   $-$0.4821(77)  &               \\
$\Phi_K$ / Hz                  &    2.98(16)     &    1.951(44)   &               \\
$\phi_J$ / Hz                  &    0.007386(90) &    0.008079(46)&               \\
$\phi_{JK}$ / Hz                 &    0.0346(33)   &    0.0117(20)  &               \\
$\phi_K$ / Hz                  &    0.404(28)    &    0.602(18)   &               \\
$L_{JK}$ / mHz           &    0.088(10)    &    0.0462(17)  &               \\
$L_{K}$ / mHz            &   $-$4.62(48)     &   $-$2.65(14)    &               \\
$l_{JK}$ / mHz           &    0.0312(55)   &       --       &               \\ \hline
$\chi_{aa}$ / MHz              &             --            &        0.0598(19)       &     0.0579(15)   \\
($\chi_{bb}-\chi_{cc}$) / MHz    &             --            &       $-$4.6868(30)       &    $-$4.688(2)     \\
$\sigma_{fit}$ / kHz           &            41.1           &          27.5           &       2.8       \\
$N_{lines}$                  &             537           &          4793           &        152       \\
$J_{min}/J_{max}$            &            6/69           &          0/69           &        0/5       \\
$K_{a,min}/K_{a,max}$        &            0/16           &          0/16           &        0/3       \\
\hline
\end{tabular}                                   
\tablenoteb{\\
$^a$\,Ground state analysis without taking the hyperfine structure into consideration. In the case of observable splitting, the central frequency is used.\\
$^b$\,Ground state analysis taking all hyperfine components into consideration, including those overlapped and the data from \cite{Halter2001} up to 14 GHz.
}
\end{table*}

The line shape of several transitions did not correspond to a single line component, but they appeared wider or with multiple peaks corresponding to the nuclear quadrupole hyperfine structure due to the two $^{14}$N nuclei. The initial fit was performed assuming that this quadrupole hyperfine structure has collapsed in the $W$ band frequency region. For the $R$ branch, the splitting became more appreciable for low $J$ values and increasing $K_a$. For the $Q$ branch, the higher $K_a$ and $J$, the larger the observed spread in the hyperfine structure. A second fit keeping in mind all hyperfine components of each transition was performed  including the 60 hyperfine components observed in \cite{Halter2001} in the 5-14 GHz range ($J$ values from 0 to 5 and $K_a$ from 0 to 2). In this fit the number of lines were drastically increased since each transition is divided into around ten hyperfine components. A total of 4793 components were included in the final fit, most of them overlapped for the same rotational transition. A comparison between the spectroscopic parameters determined in this work, with and without the nuclear quadrupole hyperfine analysis, and those found from the microwave data by \cite{Halter2001} can be found in Table\,\ref{ground_constants}. This more precise set of parameters including higher order terms (up to the octic distortion constants) reproduces within the experimental uncertainty the rotational transitions of maleonitrile in the W-band (75-110 GHz).

\begin{figure*}
\centering
\includegraphics[angle=0,width=0.78\textwidth]{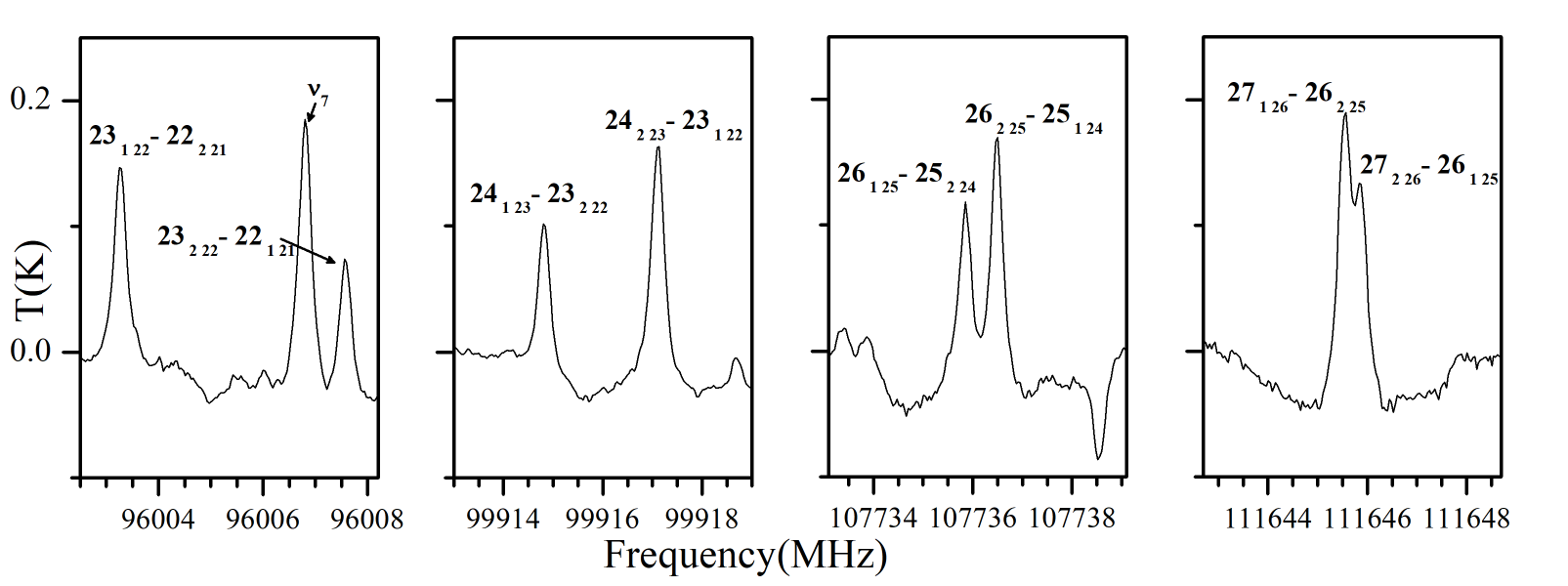}
\caption{Series of $R_{1,-1},~R_{-1,-1}$ rotational transitions of the ground state of maleonitrile, illustrating the influence of nuclear spin statistics on the transition intensities, which are related as 21-to-15 depending on the $K_{a}$+$K_{c}$ values.}
\label{Spin statistics}
\end{figure*}

\begin{figure*}
\centering
\includegraphics[angle=0,width=0.78\textwidth]{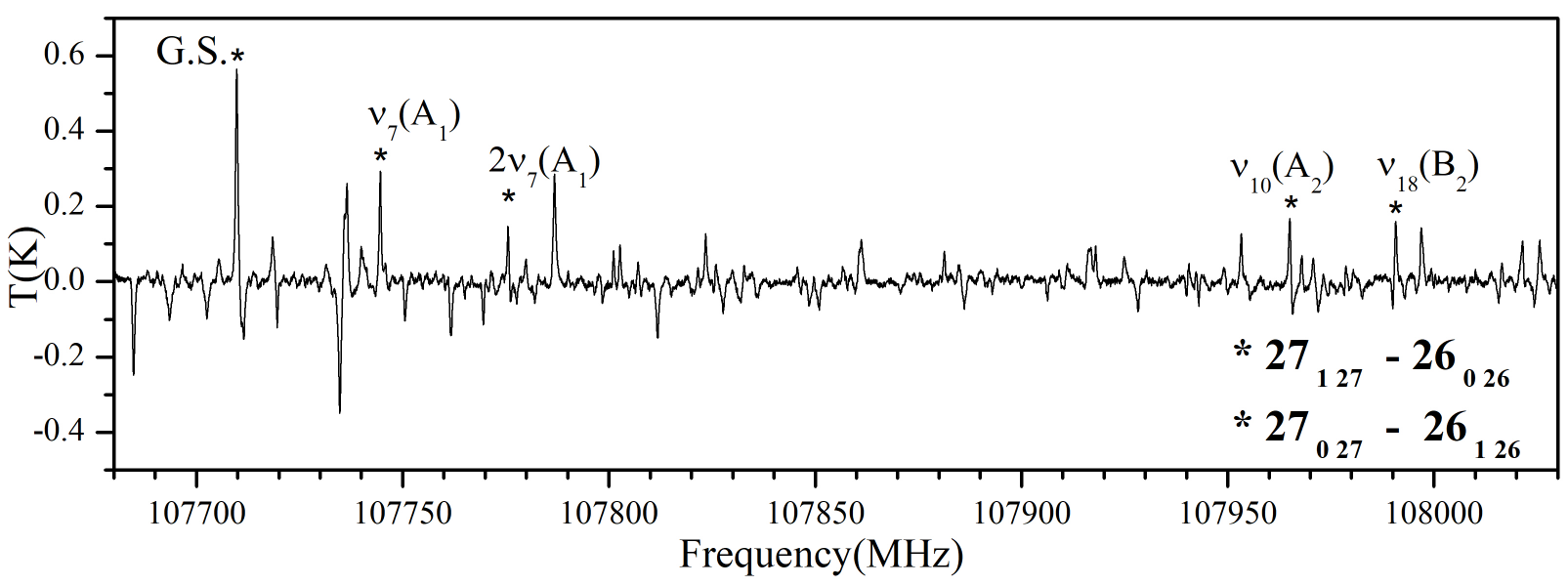}
\caption{Part of the spectrum of maleonitrile showing the satellites around a rotational transition of the ground vibrational state that correspond to the lowest vibrationally excited states.}
\label{SpecModos}
\end{figure*}

The relative intensities for the sequences of transitions alternate in function of the $J$ and the parity of $K_a$ and $K_c$, due to the nuclear spin statistics. Figure \ref{Spin statistics} illustrates a series of $R$-branch rotational transitions, highlighting the impact of nuclear spin statistics on transition intensities. Maleonitrile belongs to the group of symmetry $C_{2v}$ that makes its two nitrogen nuclei (fermions) and its two hydrogen nuclei (bosons) equivalents due to their interchange by the symmetry operators: plane $\sigma_{v}$ and axis $C_2$. The overall wave function, denoted as $\psi_{tot}$ = $\psi_{ele}\psi_{vib}\psi_{rot}\psi_{ns}$, must be antisymmetric, hence adhere to Fermi-Dirac statistics. The wave functions $\psi_{ele}$ and $\psi_{vib}$ for the ground electronic and vibrational states are symmetric, while the parity of the rotational wave function $\psi_{rot}$ depends on the $K_{a}$ and $K_{c}$ values. As detailed in \cite{Bunker1998}, the total nuclear statistical weights are (2$I_H$ + 1)$^2$(2$I_N$ + 1)$^2$ = 36, with nuclear statistical weights for rotational levels with {$K_{a} \oplus K_{c}$ = even and odd} being 15 and 21, respectively, giving raise to a ratio of 5 to 7. As observed in Fig.\,\ref{Spin statistics}, the relative intensities of all the rotational transitions within the ground vibrational state align well with the predictions made with the SPCAT program \citep{Pickett1991} considering spin statistics.

\begin{table*}
\small
\centering
\caption[]{MP2/cc-pVTZ anharmonic vibrational frequencies, first-order vibration-rotation $\alpha$ coefficients, and intensities of the fundamental modes of maleonitrile.}{
\label{vibfreq}
\begin{tabular}{ccccccc}
\hline
\hline
Mode & Energy (cm$^{-1}$) & Symmetry\,$^a$ & IR intensity (km mol$^{-1}$) & $\alpha_i$($a$) (cm$^{-1}$)\,$^b$ & $\alpha_i$($b$) (cm$^{-1}$)\,$^b$ & $\alpha_i$($c$) (cm$^{-1}$)\,$^b$ \\
\hline
1     &   3105   &  $A_1$   &   0.32  &  $-$11.3     &    9.3  &   4.1  \\
2     &   2125   &  $A_1$   &  11.01  &   26.9       &    4.0  &   4.1  \\
3     &   1602   &  $A_1$   &   0.07  &   42.4       &   $-$7.9  &  $-$1.1  \\
4     &   1209   &  $A_1$   &   0.41  &   42.8       &  $-$23.4  &  $-$8.1  \\
5     &    874   &  $A_1$   &   4.35  &  $-$30.4     &   20.2  &   8.8  \\
6     &    454   &  $A_1$   &   0.06  &  $-$48.3     &   11.7  &   3.4  \\
7     &    101   &  $A_1$   &   2.63  &  $-$31.6     &  $-$10.1  &  $-$1.8  \\
8     &    963   &  $A_2$   &   0.00  &   18.1       &   $-$2.7  &  $-$1.9  \\
9     &    588   &  $A_2$   &   0.00  &    8.0       &   $-$2.8  &  $-$1.3  \\
10    &    222   &  $A_2$   &   0.00  &   85.4       &   $-$6.6  &  $-$5.2  \\
11    &    769   &  $B_1$   &  36.04  &   13.7       &   $-$2.8  &  $-$1.5  \\
12    &    362   &  $B_1$   &   0.06  &   $-$8.0     &    2.8  &   0.0  \\
13    &   3115   &  $B_2$   &   5.96  &  $-$11.5     &    9.3  &   4.1  \\
14    &   2136   &  $B_2$   &   8.58  &   25.5       &    4.9  &   4.4  \\
15    &   1380   &  $B_2$   &   0.06  &   15.4       &   $-$7.0  &  $-$1.3  \\
16    &   1019   &  $B_2$   &   2.54  &   14.0       &   $-$2.9  &   0.7  \\
17    &    722   &  $B_2$   &   1.12  &  $-$10.3     &    1.5  &   1.7  \\
18    &    253   &  $B_2$   &   4.62  &  $-$88.0     &   $-$8.6  &  $-$4.8  \\ 
\hline
\end{tabular}
}
\tablenotec{\\
$^a$\,Symmetry species of the $C_{2v}$ point group.\\                   
$^b$\,First order vibration-rotation $\alpha$ coefficients of the $i$ vibrational mode along the axes, $a$, $b$, $c$. Those coefficients were employed to calculate the rotational constants of each vibrational mode following the expression $B_{\nu}$= $B_{e}$ $-$ $\sum_i$ $\alpha_i$ ($\nu_i$ + 1/2), where $B_{e}$ is the rotational constant in the equilibrium.}
\end{table*}

\begin{table*}
\small
\centering
\caption[]{Rotational parameters obtained from the analysis of the millimeter spectrum of maleonitrile: ground state and the lowest-energy vibrational excited states.}
{\label{vib_constants}
\begin{tabular}{lccccc}
\hline
\hline
Parameter / Unit   & Ground State\,$^a$ & $\nu_{7}$\,$^b$  & $2\nu_{7}$\,$^b$ & $\nu_{10}$\,$^b$ & \textbf{$\nu_{18}$\,$^b$} \\
\hline
$A$ / MHz              &     7389.5215(14)  &   7429.2453(16)     &  7469.036(54)      &  7312.60(15)       & 7469.656(80)    \\
$B$ / MHz              &   2672.59327(85)   &   2678.67381(88)    &  2684.4216(55)      &  2678.621(23)      & 2680.7789(11)   \\
$C$ / MHz              &   1959.52286(91)   &   1959.8650(10)     &  1960.14081(95)     &  1964.5140(22)     & 1964.3750(16)   \\
$\Delta_J$ / MHz       &      0.0032566(16) &      0.0032352(17)  &     0.00322585(70)  &     0.003217(22)   &    0.003233(14) \\
$\Delta_{JK}$ / MHz    &     $-$0.0216019(19) &     $-$0.0215016(27)  &    $-$0.022961(51)    &    $-$0.02200(80)    &   $-$0.02266(19)  \\
$\Delta_K$ / MHz       &      0.046324(22)  &      0.046870(26)   &     0.0526(16)      &     0.0438(61)     &    0.0602(30)   \\
$\delta_J$ / kHz       &      1.18701(14)   &      1.18253(14)    &    [1.18701]        &     1.169(11)      &    1.1773(75)   \\
$\delta_K$ / kHz       &      3.7917(28)    &      4.0992(44)     &    [3.7917]         &     2.27(47)       &    4.17(17)     \\
$\Phi_J$ / Hz          &      0.0176(10)    &      0.0177(10)     &                     &                    &                 \\
$\Phi_{JK}$ / Hz       &     $-$0.0532(39)    &     $-$0.0274(75)     &                     &                    &                 \\
$\Phi_{KJ}$ / Hz       &     $-$0.487(14)     &     $-$0.439(21)      &                     &                    &                 \\
$\Phi_K$ / Hz          &      2.98(16)      &      1.31(10)       &                     &                    &                 \\
$\phi_J$ / Hz          &      0.007386(90)  &      0.00737(14)    &                     &                    &                 \\
$\phi_{JK}$ / Hz         &      0.0346(33)    &      0.0232(60)     &                     &                    &                 \\
$\phi_K$ / Hz          &      0.404(28)     &      0.518(53)      &                     &                    &                 \\
$L_{JK}$ / mHz   &      0.088(10)     &                     &                     &                    &                 \\
$L_{K}$ / mHz    &     $-$4.62(48)      &                     &                     &                    &                 \\
$l_{JK}$ / mHz   &      0.0312(55)    &                     &                     &                    &                 \\
\hline
$\sigma_{fit}$ / kHz   &        41.1        &        37.8         &         37.7        &        64.5        &       49.6      \\
$N_{lines}$          &         537        &         318         &           55        &          64        &         66      \\
$J_{min} / J_{max}$    &        6/69        &        6/67         &        18/29        &       16/29        &      17/29      \\
$K_{a,min} / K_{a,max}$&        0/16        &        0/16         &          0/3        &         0/4        &        0/4      \\
$\Delta$E / cm$^{-1}$\,$^c$     &       0             &           101           &           203           &           222           &           253           \\
\hline
\end{tabular}
}
\tablenoted{\\
$^a$\,Ground state analysis without taking into consideration the hyperfine structure. In case of observable splitting, the central frequency is used.\\
$^b$\,Numbers in parentheses represent the derived uncertainty ($1\,\sigma$) of the parameter in units of the last digit.\\
$^c$\,Difference in energy between the ground state and the corresponding vibrational excited state calculated at MP2/cc-pVTZ level of theory under the anharmonic correction.
}
\end{table*}

Close to each intense rotational transition of the ground vibrational state of maleonitrile, a repeated pattern of satellites appear with an intensity that could correspond to the lowest vibrationally excited states (see Fig.\,\ref{SpecModos}). We performed quantum chemical calculations to estimate the vibrational modes under the anharmonic approximation using Gaussian 2016 \citep{Gaussian}. We employed the M\o ller Plesset second order perturbation theory (MP2) and the triple zeta correlation consistent polarized Dunning's basis set (cc-pVTZ). The calculated frequencies and infrared intensities of the vibrational modes, and the first order rovibrational coefficients $\alpha$ can be found in Table\,\ref{vibfreq}. Using the predictions of the calculated rotational constants of the lowest energy vibrationally excited state (see note on Table\,\ref{vibfreq}), we identified the four lowest-energy vibrationally excited states of maleonitrile as $\nu_{7}$($A_1$), $2\nu_{7}$($A_1$), $\nu_{10}$($A_2$), and $\nu_{18}$($B_2$). The analysis of their rotational transitions were carried out in the same manner than for the ground vibrational state. The spectroscopic parameters for these four vibrationally excited states are given in Table\,\ref{vib_constants}.

\section{Sensitive line-by-line detection of HCC$-$CH$_2$$-$CN and ($E$)-HCC$-$CH$=$CH$-$CN} \label{app:line-by-line}

\begin{figure*}
\centering
\includegraphics[angle=0,width=0.87\textwidth]{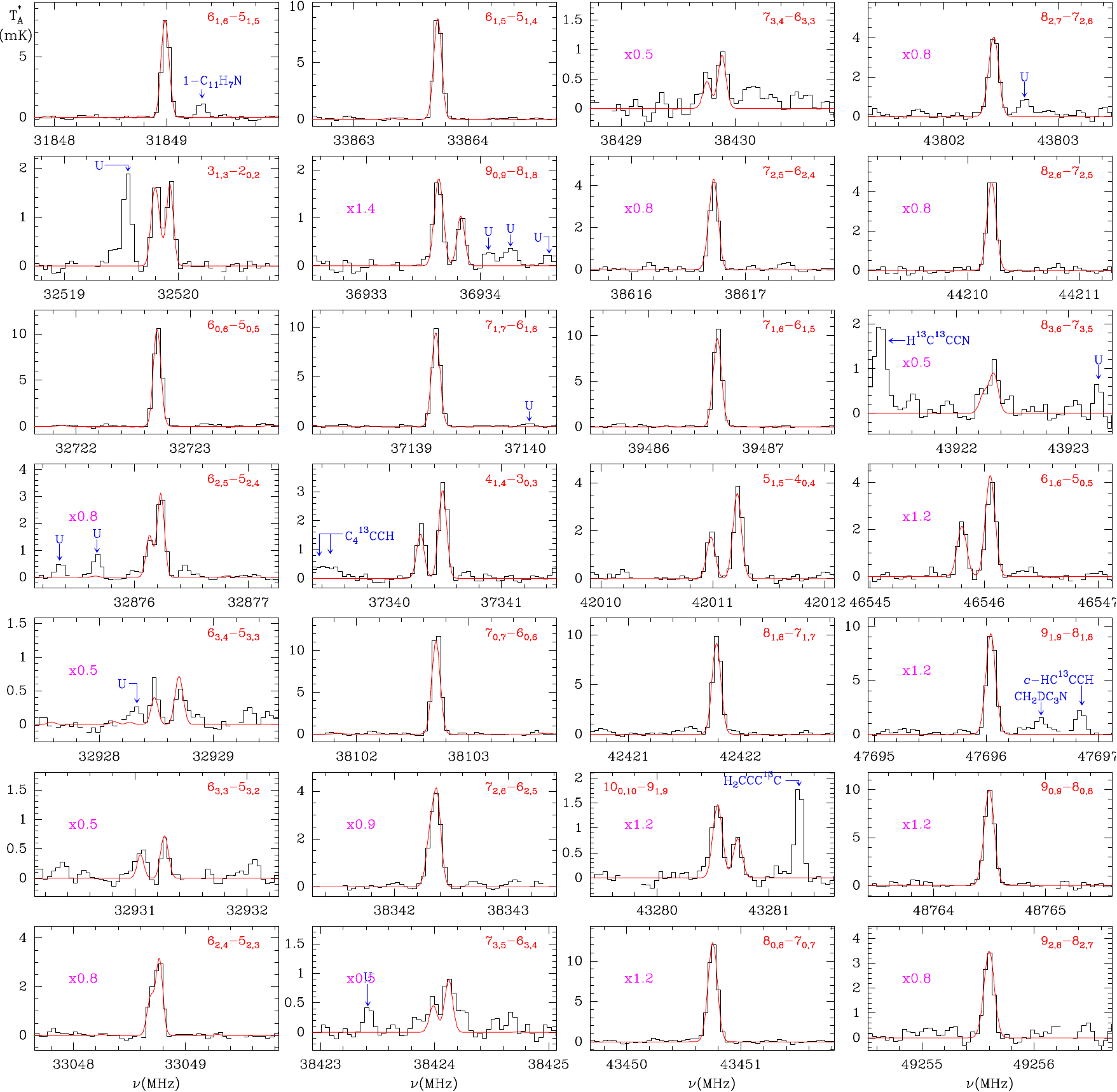}
\caption{Lines of HCC$-$CH$_2$$-$CN detected in \mbox{TMC-1}.}
\label{fig:hccch2cn}
\end{figure*}

We re-analysed the lines of HCC$-$CH$_2$$-$CN presented previously by \cite{Marcelino2021} using the latest dataset of QUIJOTE. The lines are shown in Fig.\,\ref{fig:hccch2cn}, where we also show the line profiles calculated for our best fit local thermodynamic equilibrium model using a rotational temperature 5.3 K and a column density of 1.4\,$\times$\,10$^{12}$ cm$^{-2}$. The column density is revised downwards by a factor of two with respect to the value originally determined by \cite{Marcelino2021}, 2.8\,$\times$\,10$^{12}$ cm$^{-2}$. In Fig.\,\ref{fig:hccch2cn}, the intensities of some lines have been multiplied by a scaling factor (indicated in each panel) to better reproduce the observed line intensities. The modifications of the calculated intensities are larger for the $K_a$\,=\,3 lines (a factor of two), remain low for the $K_a$\,=\,2 lines (10-40\,\%), and are minor for the $K_a$\,=\,0,1 lines. We attribute these deviations to the excitation of HCC$-$CH$_2$$-$CN, which seems to possess slightly different temperatures for intra- and inter-$K_a$ ladders.

\begin{figure*}
\centering
\includegraphics[angle=0,width=0.87\textwidth]{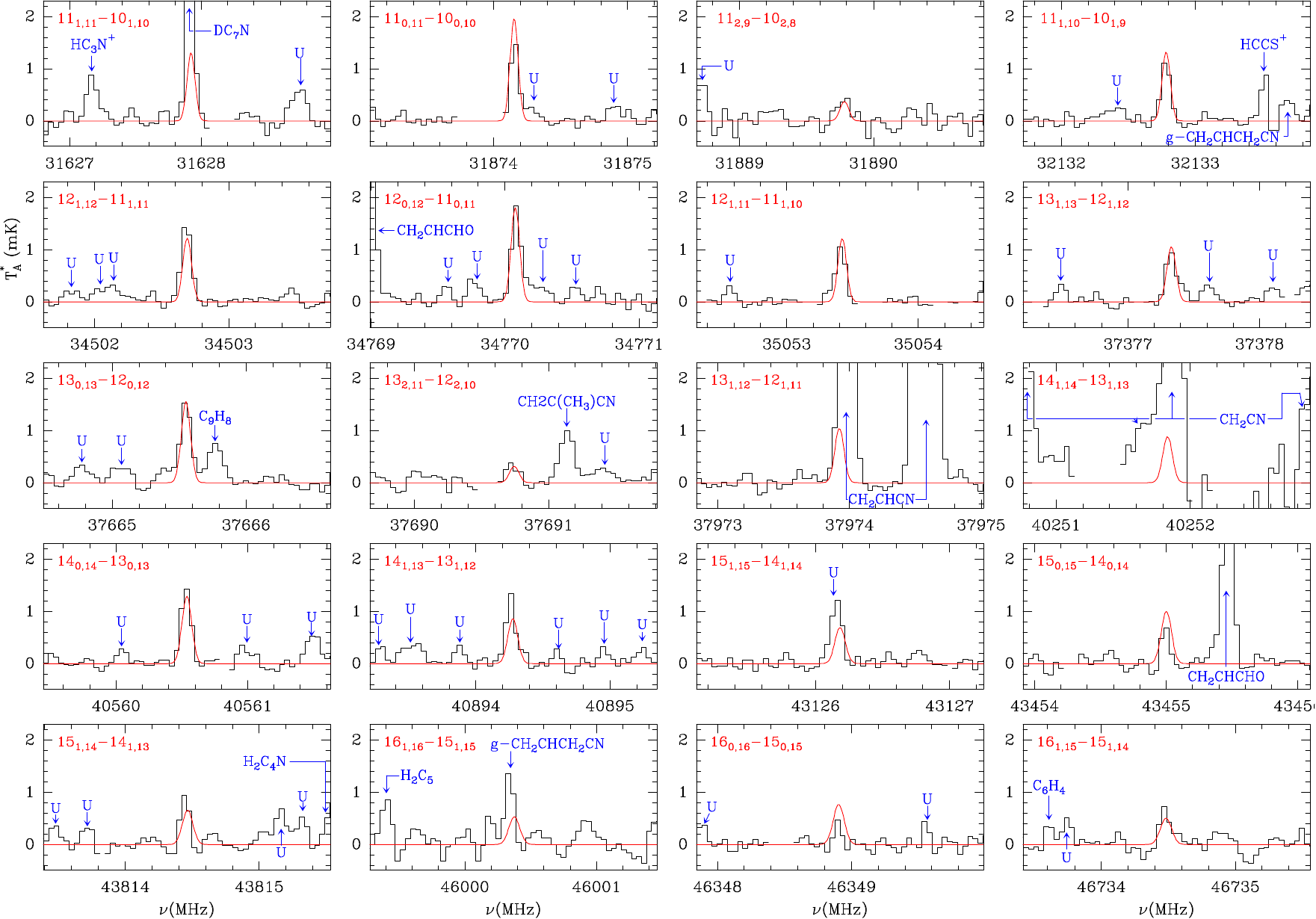}
\caption{Lines of ($E$)-HCC$-$CH$=$CH$-$CN detected in \mbox{TMC-1}.}
\label{fig:e_hccchchcn}
\end{figure*}

We also present a line-by-line detection of the $E$ isomer of HCC$-$CH$=$CH$-$CN using the latest QUIJOTE dataset. The lines are shown in Fig.\,\ref{fig:e_hccchchcn}, where we also show the calculated line profiles adopting the parameters derived from a rotation diagram, which are $T_{\rm rot}$\,=\,5.5\,$\pm$\,0.5 K and $N$\,=\,(1.7\,$\pm$\,0.2)\,$\times$\,10$^{11}$ cm$^{-2}$. This molecule had been previously detected in \mbox{TMC-1} through line stack techniques \citep{Lee2021}. The column density determined here through the line-by-line procedure is about twice lower than the value of 3\,$\times$\,10$^{11}$ cm$^{-2}$ determined by \cite{Lee2021} through line stack. As shown in Fig. \ref{fig:e_hccchchcn}, there are various intense lines that overlap with lines of ($E$)-HCC$-$CH$=$CH$-$CN, as, e.g., the 11$_{1,11}$-10$_{1,10}$ line shown in the top-left panel which is blended with a line of DC$_7$N or the 16$_{1,16}$-15$_{1,15}$ line shown at the bottom which overlaps with a line of $g$-CH$_2$CHCH$_2$CN. These overlaps indicate that line stack needs to be done with great caution.

\section{FTMW spectroscopy of CH$_2$$=$C(CN)$_2$} \label{app:ch2c_cn_2}

The rotational spectrum of 1,1-dicyanoethylene was observed using a Balle-Flygare narrowband-type Fourier-transform microwave (FTMW) spectrometer operating in the frequency region of 4-40 GHz \citep{Endo1994,Cabezas2016}. The species CH$_2$$=$C(CN)$_2$ was produced in a supersonic expansion by a pulsed electric discharge of a gas mixture of 0.3\,\% vinyl cyanide (CH$_2$CHCN) diluted in argon. The gas mixture was flowed through a pulsed-solenoid valve that is accommodated in the backside of one of the cavity mirrors and aligned parallel to the optical axis of the resonator. A pulse voltage of 1500 V with a duration of 450 $\mu$s was applied between stainless-steel electrodes attached to the exit of the pulsed discharge nozzle, resulting in an electric discharge synchronized with the gas expansion. The resulting products generated in the discharge were then probed by FTMW spectroscopy capable of resolving small hyperfine splittings.

The frequency predictions were done using the rotational constants derived by \citet{Tan1978}. A total of 15 $b$-type rotational transitions including $R$ and $Q$ branches were observed in the 10-26 GHz region.
The final dataset consisted of 129 hyperfine components. Table\,\ref{tab_lab_h2cccn2} contains the experimental frequencies for all the observed hyperfine components. Rotational, centrifugal, and nuclear quadrupole coupling constants were determined by fitting the transition frequencies with the SPFIT program \citep{Pickett1991} to a Watson's $A$-reduced Hamiltonian for asymmetric top molecules, with the following form: $H$ = $H_R$ + $H_Q$ \citep{Watson1977}, where $H_R$ contains the rotational and centrifugal distortion parameters, and $H_Q$ contains the quadrupole coupling interactions. The energy levels involved in each transition are labeled with the quantum numbers $J$, $K_{a}$, $K_{c}$, $I,$ and $F$. Since 1,1-dicyanoethylene has two equivalent $^{14}$N nuclei with a spin of $I_N$ = 1,
the following coupling scheme was used \textbf{F} = \textbf{J} + \textbf{I}, where \textbf{I} = 0, 1, 2. The analysis rendered a set of experimental spectroscopic parameters, which are compared with those derived by \cite{Tan1978} in Table\,\ref{ctes_h2cccn2}.

\begin{table}
\small
\centering
\caption{Spectroscopic parameters of CH$_2$C$=$C(CN)$_2$.}
\label{ctes_h2cccn2}
\centering
\begin{tabular}{{lccc}}
\hline
\hline
Parameter / Unit         & This work        & \citet{Tan1978}  \\
\hline
$A$ / MHz                   &     6579.26967(89)\,$^a$&           6579.266(5)  \\
$B$ / MHz                   &   2882.033545(304)    &           2882.032(2)  \\
$C$ / MHz                   &     2000.92392(32)    &           2000.925(2)  \\
$\Delta_J$ / kHz            &         1.2593(63)    &            1.2506(63)  \\
$\Delta_{JK}$ / kHz         &         -8.584(69)    &          -8.5636(252)  \\
$\Delta_K$ / kHz            &        32.315(136)    &          32.2825(490)  \\
$\delta_J$ / kHz            &       0.55633(226)    &            0.5568(21)  \\
$\delta_K$ / kHz            &         1.364(131)    &           1.3311(335)  \\
$\chi_{aa}$ / MHz           &      -2.64535(226)    &            -           \\
$\chi_{bb}$ / MHz           &       0.55546(175)    &            -           \\
$N_{lines}$               &           129\,$^b$     &             59\,$^c$     \\
$\sigma$ / kHz             &           2.9         &             45         \\
\hline
\hline
\end{tabular}
\tablenotee{\\
$^a$\,The uncertainties (in parentheses) are in units of the last
significant digits. $^b$\,Number of hyperfine components. $^c$\,Number of pure rotational transitions.
}
\end{table}
\normalsize

\small
\begin{longtable}{cccccccccccr}
\caption[]{Laboratory-observed transition frequencies for CH$_2$$=$C(CN)$_2$.}
\label{tab_lab_h2cccn2} \\
\hline
\hline
 $J'$ & $K'_a$ & $K'_c$ & $I'$ & $F'$  & $J''$ & $K''_a$ & $K''_c$ & $I''$& $F''$ & $\nu_{obs}$  &  Obs-Calc \\
      &        &        &        &       &       &         &         &        &     &   (MHz)      &   (MHz)      \\
\hline
\endfirsthead
\caption{Continued.}\\
\hline
\hline
 $J'$ & $K'_a$ & $K'_c$ & $I'$ & $F'$  & $J''$ & $K''_a$ & $K''_c$ & $I''$& $F''$ & $\nu_{obs}$  &  Obs-Calc \\
      &        &        &        &       &       &         &         &        &     &   (MHz)      &   (MHz)      \\
\hline
\endhead
\hline
\endfoot
\hline
\endlastfoot
\hline
4 & 1 & 3 & 0 & 4 & 4 & 0 & 4 & 0 & 4 &   10024.193  &   0.005 \\
4 & 1 & 3 & 1 & 4 & 4 & 0 & 4 & 1 & 4 &   10024.216  &  $-$0.001 \\
4 & 1 & 3 & 2 & 6 & 4 & 0 & 4 & 2 & 6 &   10024.317  &  $-$0.003 \\
4 & 1 & 3 & 1 & 5 & 4 & 0 & 4 & 1 & 5 &   10024.732  &   0.000 \\
4 & 1 & 3 & 1 & 3 & 4 & 0 & 4 & 1 & 3 &   10024.864  &   0.000 \\
4 & 1 & 3 & 2 & 5 & 4 & 0 & 4 & 2 & 5 &   10024.936  &  $-$0.001 \\
4 & 2 & 2 & 2 & 2 & 4 & 1 & 3 & 2 & 2 &   10218.562  &   0.002 \\
4 & 2 & 2 & 1 & 4 & 4 & 1 & 3 & 1 & 4 &   10218.662  &  $-$0.000 \\
4 & 2 & 2 & 2 & 6 & 4 & 1 & 3 & 2 & 6 &   10218.724  &  $-$0.004 \\
4 & 2 & 2 & 2 & 3 & 4 & 1 & 3 & 2 & 3 &   10218.868  &  $-$0.000 \\
4 & 2 & 2 & 1 & 5 & 4 & 1 & 3 & 1 & 5 &   10218.989  &  $-$0.000 \\
4 & 2 & 2 & 2 & 2 & 4 & 1 & 3 & 2 & 3 &   10219.075  &   0.002 \\
4 & 2 & 2 & 1 & 3 & 4 & 1 & 3 & 1 & 3 &   10219.075  &   0.002 \\
3 & 0 & 3 & 2 & 3 & 2 & 1 & 2 & 2 & 2 &   11156.173  &   0.003 \\
3 & 0 & 3 & 2 & 2 & 2 & 1 & 2 & 0 & 2 &   11156.237  &   0.002 \\
3 & 0 & 3 & 2 & 4 & 2 & 1 & 2 & 2 & 3 &   11156.492  &  $-$0.002 \\
3 & 0 & 3 & 1 & 4 & 2 & 1 & 2 & 1 & 3 &   11156.533  &  $-$0.000 \\
3 & 0 & 3 & 1 & 2 & 2 & 1 & 2 & 1 & 1 &   11156.563  &  $-$0.002 \\
3 & 0 & 3 & 0 & 3 & 2 & 1 & 2 & 0 & 2 &   11156.776  &   0.001 \\
3 & 0 & 3 & 2 & 5 & 2 & 1 & 2 & 2 & 4 &   11156.816  &  $-$0.000 \\
3 & 0 & 3 & 1 & 3 & 2 & 1 & 2 & 1 & 2 &   11156.839  &   0.002 \\
2 & 1 & 2 & 2 & 1 & 1 & 0 & 1 & 2 & 1 &   12580.837  &  $-$0.000 \\
2 & 1 & 2 & 0 & 2 & 1 & 0 & 1 & 2 & 1 &   12581.001  &  $-$0.000 \\
2 & 1 & 2 & 2 & 3 & 1 & 0 & 1 & 2 & 3 &   12581.185  &  $-$0.001 \\
2 & 1 & 2 & 1 & 2 & 1 & 0 & 1 & 1 & 1 &   12581.912  &   0.002 \\
2 & 1 & 2 & 2 & 2 & 1 & 0 & 1 & 0 & 1 &   12581.945  &  $-$0.001 \\
2 & 1 & 2 & 2 & 4 & 1 & 0 & 1 & 2 & 3 &   12582.083  &   0.000 \\
2 & 1 & 2 & 2 & 3 & 1 & 0 & 1 & 2 & 2 &   12582.381  &   0.003 \\
2 & 2 & 1 & 0 & 2 & 2 & 1 & 2 & 0 & 2 &   13733.125  &   0.000 \\
2 & 2 & 1 & 2 & 1 & 2 & 1 & 2 & 2 & 1 &   13733.498  &   0.003 \\
2 & 2 & 1 & 1 & 2 & 2 & 1 & 2 & 1 & 2 &   13733.498  &   0.003 \\
2 & 2 & 1 & 2 & 4 & 2 & 1 & 2 & 2 & 4 &   13733.998  &  $-$0.004 \\
3 & 2 & 2 & 1 & 4 & 3 & 1 & 3 & 1 & 3 &   15136.991  &   0.002 \\
3 & 2 & 2 & 1 & 2 & 3 & 1 & 3 & 1 & 3 &   15136.991  &   0.002 \\
3 & 2 & 2 & 1 & 3 & 3 & 1 & 3 & 1 & 3 &   15136.991  &   0.002 \\
3 & 2 & 2 & 2 & 5 & 3 & 1 & 3 & 2 & 5 &   15137.222  &   0.001 \\
3 & 2 & 2 & 2 & 4 & 3 & 1 & 3 & 2 & 5 &   15137.222  &   0.001 \\
3 & 2 & 2 & 1 & 3 & 3 & 1 & 3 & 1 & 4 &   15137.915  &  $-$0.001 \\
3 & 2 & 2 & 1 & 4 & 3 & 1 & 3 & 1 & 4 &   15137.915  &  $-$0.001 \\
3 & 2 & 2 & 1 & 3 & 3 & 1 & 3 & 1 & 2 &   15138.241  &  $-$0.000 \\
3 & 2 & 2 & 1 & 2 & 3 & 1 & 3 & 1 & 2 &   15138.241  &  $-$0.000 \\
3 & 2 & 2 & 2 & 3 & 3 & 1 & 3 & 2 & 4 &   15138.381  &   0.000 \\
3 & 2 & 2 & 2 & 5 & 3 & 1 & 3 & 2 & 4 &   15138.381  &   0.000 \\
3 & 2 & 2 & 0 & 3 & 3 & 1 & 3 & 2 & 4 &   15138.381  &   0.000 \\
3 & 2 & 2 & 2 & 4 & 3 & 1 & 3 & 2 & 4 &   15138.381  &   0.000 \\
3 & 1 & 3 & 0 & 3 & 2 & 0 & 2 & 0 & 2 &   16201.534  &  $-$0.000 \\
3 & 1 & 3 & 2 & 3 & 2 & 0 & 2 & 2 & 2 &   16201.547  &   0.000 \\
3 & 1 & 3 & 1 & 4 & 2 & 0 & 2 & 1 & 3 &   16201.612  &  $-$0.001 \\
3 & 1 & 3 & 1 & 3 & 2 & 0 & 2 & 1 & 2 &   16201.650  &  $-$0.003 \\
3 & 1 & 3 & 2 & 5 & 2 & 0 & 2 & 2 & 4 &   16201.715  &  $-$0.002 \\
3 & 1 & 3 & 2 & 4 & 2 & 0 & 2 & 2 & 3 &   16201.744  &   0.003 \\
3 & 1 & 3 & 1 & 2 & 2 & 0 & 2 & 1 & 1 &   16201.778  &  $-$0.003 \\
4 & 0 & 4 & 2 & 4 & 3 & 1 & 3 & 2 & 3 &   16216.684  &   0.000 \\
4 & 0 & 4 & 2 & 3 & 3 & 1 & 3 & 2 & 2 &   16216.703  &   0.000 \\
4 & 0 & 4 & 1 & 5 & 3 & 1 & 3 & 1 & 4 &   16216.823  &   0.001 \\
4 & 0 & 4 & 2 & 5 & 3 & 1 & 3 & 2 & 4 &   16216.860  &  $-$0.003 \\
4 & 0 & 4 & 2 & 2 & 3 & 1 & 3 & 2 & 1 &   16216.860  &  $-$0.003 \\
4 & 0 & 4 & 1 & 3 & 3 & 1 & 3 & 1 & 2 &   16216.874  &  $-$0.000 \\
4 & 0 & 4 & 0 & 4 & 3 & 1 & 3 & 0 & 3 &   16216.919  &   0.004 \\
4 & 0 & 4 & 1 & 4 & 3 & 1 & 3 & 1 & 3 &   16216.949  &  $-$0.000 \\
4 & 0 & 4 & 2 & 6 & 3 & 1 & 3 & 2 & 5 &   16216.970  &  $-$0.000 \\
4 & 1 & 4 & 2 & 3 & 3 & 0 & 3 & 2 & 2 &   19616.717  &   0.000 \\
4 & 1 & 4 & 2 & 4 & 3 & 0 & 3 & 2 & 3 &   19616.778  &  $-$0.000 \\
4 & 1 & 4 & 1 & 5 & 3 & 0 & 3 & 1 & 4 &   19616.863  &   0.001 \\
4 & 1 & 4 & 0 & 4 & 3 & 0 & 3 & 0 & 3 &   19616.896  &  $-$0.005 \\
4 & 1 & 4 & 2 & 5 & 3 & 0 & 3 & 2 & 4 &   19616.925  &  $-$0.000 \\
4 & 1 & 4 & 1 & 4 & 3 & 0 & 3 & 1 & 3 &   19616.949  &   0.008 \\
4 & 1 & 4 & 2 & 6 & 3 & 0 & 3 & 2 & 5 &   19616.973  &  $-$0.000 \\
5 & 0 & 5 & 2 & 4 & 4 & 1 & 4 & 2 & 3 &   20984.408  &  $-$0.003 \\
5 & 0 & 5 & 2 & 5 & 4 & 1 & 4 & 2 & 4 &   20984.423  &   0.000 \\
5 & 0 & 5 & 1 & 6 & 4 & 1 & 4 & 1 & 5 &   20984.477  &  $-$0.012 \\
5 & 0 & 5 & 2 & 3 & 4 & 1 & 4 & 2 & 2 &   20984.505  &   0.002 \\
5 & 0 & 5 & 2 & 6 & 4 & 1 & 4 & 2 & 5 &   20984.538  &   0.002 \\
5 & 0 & 5 & 0 & 5 & 4 & 1 & 4 & 0 & 4 &   20984.542  &  $-$0.001 \\
5 & 0 & 5 & 2 & 7 & 4 & 1 & 4 & 2 & 6 &   20984.586  &   0.000 \\
2 & 2 & 0 & 1 & 2 & 1 & 1 & 1 & 1 & 2 &   22758.322  &   0.001 \\
2 & 2 & 0 & 1 & 2 & 1 & 1 & 1 & 1 & 1 &   22758.487  &  $-$0.000 \\
2 & 2 & 0 & 0 & 2 & 1 & 1 & 1 & 2 & 1 &   22758.494  &  $-$0.002 \\
2 & 2 & 0 & 2 & 4 & 1 & 1 & 1 & 2 & 3 &   22758.701  &   0.001 \\
2 & 2 & 0 & 1 & 3 & 1 & 1 & 1 & 1 & 2 &   22759.208  &   0.000 \\
2 & 2 & 0 & 1 & 1 & 1 & 1 & 1 & 1 & 0 &   22759.453  &   0.001 \\
2 & 2 & 0 & 2 & 3 & 1 & 1 & 1 & 2 & 2 &   22759.635  &   0.001 \\
2 & 2 & 0 & 1 & 1 & 1 & 1 & 1 & 1 & 1 &   22759.864  &  $-$0.004 \\
2 & 2 & 0 & 2 & 3 & 1 & 1 & 1 & 2 & 3 &   22759.885  &   0.001 \\
2 & 2 & 0 & 2 & 2 & 1 & 1 & 1 & 0 & 1 &   22760.073  &   0.001 \\
5 & 1 & 5 & 2 & 5 & 4 & 0 & 4 & 2 & 4 &   23039.300  &  $-$0.000 \\
5 & 1 & 5 & 1 & 6 & 4 & 0 & 4 & 1 & 5 &   23039.372  &  $-$0.000 \\
5 & 1 & 5 & 2 & 3 & 4 & 0 & 4 & 2 & 2 &   23039.394  &   0.002 \\
5 & 1 & 5 & 1 & 4 & 4 & 0 & 4 & 1 & 3 &   23039.412  &  $-$0.003 \\
5 & 1 & 5 & 2 & 6 & 4 & 0 & 4 & 2 & 5 &   23039.412  &  $-$0.003 \\
5 & 1 & 5 & 1 & 5 & 4 & 0 & 4 & 1 & 4 &   23039.447  &  $-$0.003 \\
5 & 1 & 5 & 2 & 7 & 4 & 0 & 4 & 2 & 6 &   23039.479  &   0.006 \\
6 & 1 & 5 & 2 & 6 & 5 & 2 & 4 & 2 & 5 &   23883.430  &  $-$0.004 \\
6 & 1 & 5 & 2 & 5 & 5 & 2 & 4 & 2 & 4 &   23883.579  &   0.002 \\
6 & 1 & 5 & 1 & 5 & 5 & 2 & 4 & 1 & 4 &   23883.579  &   0.002 \\
6 & 1 & 5 & 1 & 7 & 5 & 2 & 4 & 1 & 6 &   23883.579  &   0.002 \\
6 & 1 & 5 & 2 & 8 & 5 & 2 & 4 & 2 & 7 &   23883.715  &   0.003 \\
6 & 1 & 5 & 0 & 6 & 5 & 2 & 4 & 0 & 5 &   23883.715  &   0.003 \\
6 & 1 & 5 & 1 & 6 & 5 & 2 & 4 & 1 & 5 &   23883.715  &   0.003 \\
6 & 1 & 5 & 2 & 4 & 5 & 2 & 4 & 2 & 3 &   23883.715  &   0.003 \\
6 & 0 & 6 & 1 & 7 & 5 & 1 & 5 & 1 & 6 &   25467.808  &  $-$0.000 \\
6 & 0 & 6 & 1 & 5 & 5 & 1 & 5 & 1 & 4 &   25467.834  &  $-$0.005 \\
6 & 0 & 6 & 2 & 7 & 5 & 1 & 5 & 2 & 6 &   25467.847  &  $-$0.000 \\
6 & 0 & 6 & 1 & 6 & 5 & 1 & 5 & 1 & 5 &   25467.864  &   0.004 \\
6 & 0 & 6 & 2 & 8 & 5 & 1 & 5 & 2 & 7 &   25467.883  &   0.003 \\
3 & 2 & 2 & 2 & 3 & 2 & 1 & 1 & 0 & 2 &   25740.168  &   0.001 \\
3 & 2 & 2 & 0 & 3 & 2 & 1 & 1 & 0 & 2 &   25740.168  &   0.001 \\
3 & 2 & 2 & 2 & 1 & 2 & 1 & 1 & 0 & 2 &   25740.168  &   0.001 \\
3 & 2 & 2 & 2 & 2 & 2 & 1 & 1 & 0 & 2 &   25740.168  &   0.001 \\
3 & 2 & 2 & 2 & 1 & 2 & 1 & 1 & 2 & 1 &   25740.210  &   0.000 \\
3 & 2 & 2 & 2 & 2 & 2 & 1 & 1 & 2 & 1 &   25740.210  &   0.000 \\
3 & 2 & 2 & 1 & 3 & 2 & 1 & 1 & 1 & 2 &   25740.210  &   0.000 \\
3 & 2 & 2 & 1 & 2 & 2 & 1 & 1 & 1 & 2 &   25740.210  &   0.000 \\
3 & 2 & 2 & 2 & 3 & 2 & 1 & 1 & 2 & 4 &   25740.269  &  $-$0.000 \\
3 & 2 & 2 & 2 & 5 & 2 & 1 & 1 & 2 & 4 &   25740.269  &  $-$0.000 \\
3 & 2 & 2 & 2 & 4 & 2 & 1 & 1 & 2 & 4 &   25740.269  &  $-$0.000 \\
3 & 2 & 2 & 0 & 3 & 2 & 1 & 1 & 2 & 4 &   25740.269  &  $-$0.000 \\
3 & 2 & 2 & 1 & 4 & 2 & 1 & 1 & 1 & 3 &   25740.386  &  $-$0.002 \\
3 & 2 & 2 & 1 & 2 & 2 & 1 & 1 & 1 & 3 &   25740.386  &  $-$0.002 \\
3 & 2 & 2 & 1 & 3 & 2 & 1 & 1 & 1 & 3 &   25740.386  &  $-$0.002 \\
3 & 2 & 2 & 1 & 2 & 2 & 1 & 1 & 1 & 1 &   25740.485  &  $-$0.002 \\
3 & 2 & 2 & 2 & 3 & 2 & 1 & 1 & 2 & 3 &   25740.514  &   0.006 \\
3 & 2 & 2 & 2 & 4 & 2 & 1 & 1 & 2 & 3 &   25740.514  &   0.006 \\
3 & 2 & 2 & 0 & 3 & 2 & 1 & 1 & 2 & 3 &   25740.514  &   0.006 \\
3 & 2 & 2 & 2 & 2 & 2 & 1 & 1 & 2 & 3 &   25740.514  &   0.006 \\
3 & 2 & 2 & 2 & 3 & 2 & 1 & 1 & 2 & 2 &   25740.587  &  $-$0.003 \\
3 & 2 & 2 & 0 & 3 & 2 & 1 & 1 & 2 & 2 &   25740.587  &  $-$0.003 \\
3 & 2 & 2 & 2 & 1 & 2 & 1 & 1 & 2 & 2 &   25740.587  &  $-$0.003 \\
3 & 2 & 2 & 2 & 2 & 2 & 1 & 1 & 2 & 2 &   25740.587  &  $-$0.003 \\
\hline
\end{longtable}

\end{appendix}

\end{document}